\documentclass[aps,a4paper]{revtex4}
\usepackage{graphicx}

\newcommand{\be}{\begin{equation}}
\newcommand{\ee}{\end{equation}}
\newcommand{\ba}{\begin{array}}
\newcommand{\ea}{\end{array}}
\newcommand{\bea}{\begin{eqnarray}}
\newcommand{\eea}{\end{eqnarray}}
\begin{document}

\title{ Exact Expressions for Minor Hysteresis Loops in the Random Field
Ising Model on a Bethe Lattice at Zero Temperature }

\author{Prabodh Shukla}
\email{shukla@nehu.ac.in}
\affiliation{Physics Department \\ North Eastern Hill
University \\Shillong-793 022,India.} 

\begin{abstract}

We obtain exact expressions for the minor hysteresis loops in the
ferromagnetic random field Ising model on a Bethe lattice at zero
temperature in the case when the driving field is cycled infinitely
slowly. 

\end{abstract}

\maketitle

\section{Introduction}

Hysteresis is a common phenomenon but uncommonly difficult to treat
analytically. Theoretically, hysteresis is expected to vanish as the
frequency of the driving field goes to zero, or its period goes to
infinity. However, in many cases hysteresis persists over the longest
experimental time scales. For example, there are indications that
hysteresis would be observed in a permanent magnet even if the applied
field were to be cycled so slowly as to take the entire life of an
experimentalist to complete one loop.  It is of practical importance to
make a theory for this kind of hysteresis which persists over the longest
practical time scales, and where the effect of temperature on the area and
the shape of the hysteresis loop is small. A reasonable starting point for
the theory may be the Glauber dynamics of the Ising model at temperature
$T$, driven by a field of frequency $\omega$. However, this is difficult
to treat analytically. A simpler version of this model which appears to be
adequate for our purpose is obtained by taking the limit $T=0$ first, and
then the limit $\omega=0$.  The zero-temperature, zero-frequency version
produces realistic looking hysteresis loops if one incorporates a Gaussian
distribution of on-site quenched random field in the model.  Thus the
zero-temperature single-spin-flip dynamics of the ferromagnetic random
field Ising model (RFIM) was proposed as a model of hysteresis and
Barkhausen noise by Sethna et. al.~\cite{sethna}.  Anti-ferromagnetic RFIM
is also interesting in the context of relaxation without Barkhausen
noise~\cite{shukla1}.  The difficulty in the analytical treatment of the
models comes from the presence of the quenched random field. The
zero-temperature dynamics of RFIM can not be solved exactly in two or
three dimensions (so far). We have obtained the major hysteresis loop for
the ferromagnetic RFIM in one dimension~\cite{shukla2} and on a Bethe
lattice~\cite{shukla3}. The Barkhausen noise on the major loop has been
analyzed~\cite{shukla4}. Minor hysteresis loops in the ferromagnetic RFIM
have been obtained in one dimension~\cite{shukla5}. Anti-ferromagnetic
RFIM is apparently more difficult, and its analytic solution is limited so
far to the major hysteresis loop in one dimension for a rectangular
distribution of the quenched field of width $2 \Delta$ where $\Delta \le
|J|$, $J$ being the strength of the nearest neighbor interaction. 

In the following, we present the solution of minor loops in the
ferromagnetic RFIM on a Bethe lattice. This is an extension of the work
presented in reference~\cite{shukla5}, and the completion of a problem
which remained open in reference~\cite{shukla3}; therefore the reader is
assumed to be familiar with these references. Several important aspects of
the analysis and simulations of the model which are discussed in
references ~\cite{shukla3} and ~\cite{shukla5} are not repeated here in
order to save space, or mentioned only briefly for the sake of
completeness. It was shown in one dimension~\cite{shukla5} that a reversal
of the applied field by an amount $2J$ from anywhere on the major loop
brings the system on the opposite half of the major loop. The reversed
trajectory merges with the opposite half of the major loop for the portion
of the reversed field exceeding $2J$. Thus, the width (along the applied
field axis)  of a minor loop which touches both halves of the major loop
(but does not merge with either of them) is constant and equal to $2J$
irrespective of the position of the minor loop. The shape of the minor
loop does depend upon its position inside the major loop.  This physically
interesting result is now shown to hold on all Bethe lattices irrespective
of the coordination number $z$ of the lattice. 

\section{Major hysteresis loop}

In this section, we recall the model briefly, and the solution of the
major hysteresis loop obtained earlier~\cite{shukla3}.  The RFIM in an
external field $h_{ext}$ is characterized by the Hamiltonian:

\be H=-J \sum_{i,j} s_{i} s_{j} - \sum_{i} h_{i} s_{i} - h_{ext} \sum_{i}
s_{i} \ee

The sum in the first term is restricted to pairs of nearest neighbors on a
Bethe lattice of coordination number $z$. The external field $h_{ext}$ is
cycled from $-\infty$ to $+\infty$ and back to $-\infty$. This takes the
system around its major hysteresis loop. Spins turn up on the lower half
of the loop, and turn down again on the upper half. The applied field
changes very slowly. Equivalently, at each value of the external field,
the system is allowed adequate time to attain a relaxed state with each
spin pointing along the net field at its site. In the relaxed state at
$h_{ext}=h$, the probability that an arbitrary site $i$ is up is given by,

\be p(h)=\sum_{n=0}^{z} \left( \ba{c} {z} \\ {n} \ea \right) 
[P^{*}(h)]^{n} [1-P^{*}(h)]^{z-n} p_{n}(h)  \ee

Here $P^{*}(h)$ is the conditional probability that a nearest neighbor of
site $i$ is up before site $i$ is relaxed , and $p_{n}(h)$ is the
probability that the site $i$ with quenched field $h_{i}$ can turn up in
applied field $h$ if $n$ of its nearest neighbors are up. The starting
state on the lower half of the major loop has all sites down, and the
starting state on the upper half has all sites up. Thus, on the lower half
of the major loop, $P^{*}(h)$ denotes the conditional probability that a
nearest neighbor of site $i$ turns up before site $i$. On the upper half,
$P^{*}(h)$ denotes the conditional probability that a nearest neighbor of
site $i$ turns down after site $i$. We distinguish the two situations by
putting a subscript $l$ for the lower half and $u$ for the upper half.
This gives

\be P^{*}_{l}(h)=\sum_{n=0}^{z-1} \left( \ba{c} {z-1} \\ {n} \ea \right)
[P^{*}_{l}(h)]^{n} [1-P^{*}_{l}(h)]^{z-1-n} p_{n}(h), \ee

and

\be P^{*}_{u}(h)=\sum_{n=0}^{z-1} \left( \ba{c} {z-1} \\ {n} \ea \right)
[P^{*}_{u}(h)]^{n} [1-P^{*}_{u}(h)]^{z-1-n} p_{n+1}(h)  \ee

We note that $p_{n+1}(h)=p_{n}(h+2J)$, and therefore $P^{*}_{u}(h)  =
P^{*}_{l}(h+2J)$. Here, $P^{*}_{u}(h)$ is the conditional probability that
given a site which is up on the upper half of the major loop at field $h$,
its nearest neighbor is also up. Similarly $P^{*}_{l}(h+2J)$ is the
conditional probability that given a site which is down on the lower half
of the major loop at field $h+2J$, its nearest neighbor is up. 

\section{Starting a Minor Loop}

Suppose we are on the lower part of the major loop when the applied field
is reversed from $h$ to $h^{\prime}$ ($h^{\prime} \le h$) to generate the
upper half of a minor hysteresis loop. We ask the question, what is the
probability that an arbitrary site $i$ which was up at $h$ turns down at
$h^{\prime}$. In order to compute this probability correctly, we must take
into account the irreversibility of the zero-temperature dynamics. 
Consider a site $i$ which is down on the lower half of the major loop at
an applied field $h-\delta h$ but turns up at $h$, where $\delta h$ is an
arbitrarily small field. Once site $i$ has turned up, it may not turn down
immediately if the field is rolled back to the value $h - \delta h$. The
reason is as follows. When site $i$ turns up, it increases the net field
on each of its nearest neighbors by an amount $2J$. The increased field
may cause one or more nearest neighbors of site $i$ to turn up. Suppose
$n_{a}$ nearest neighbors of site $i$ were already up before site $i$
turned up, and $n_{b}$ nearest neighbors turn up after site $i$ turns up.
Clearly, $n_{b}$ must lie in the range $ 0 \le n_{b} \le z-n_{a}$. The
$n_{b}$ neighbors increase the local field at site $i$ by a finite amount
$2 n_{b} J$. Therefore an infinitesimal decease in the applied field will
not cause site $i$ to turn down unless $n_{b}=0$. A site $i$ with $n_{b} >
0$ will turn down in decreasing applied field only after all of the
$n_{b}$ nearest neighbors have turned down. When the field is reversed to
$h^{\prime} < h$, none of the $n_{a}$ neighbors (which turned up before
site $i$ turned up) could possibly turn down before site $i$ turns down. 
This leaves the other $n_{b}$ neighbors which turned up after site $i$. 
The $n_{b}$ neighbors turned up because the field on them increased by an
amount $2J$ after site $i$ turned up. In decreasing field $h^{\prime}$,
the $n_{b}$ neighbors will turn down before site $i$ turns down. At
$h^{\prime}=h-2J$, all of the $n_{b}$ neighbors would have turned down
with the result that the nearest neighbors of a site $i$ which are up at
$h^{\prime}=h-2J$ are precisely those which were up before site $i$
flipped up. These neighbors will turn down on the reverse trajectory only
after site $i$ turns down. Thus, given an up site $i$ at $h^{\prime}=h-2J$
on the upper half of the minor loop, the conditional probability
$P^{*}_{u}(h-2J)$ that a nearest neighbor of $i$ is up is equal to
$P^{*}_{l}(h)$, where $P^{*}_{l}(h)$ is the conditional probability that a
nearest neighbor of site $i$ is up at field h given that site $i$ is down
on the lower half of the major loop at field $h$.  The probability that
the site $i$ is up at $h-2J$ is given by the equation,

\be p(h-2J)=\sum_{n=0}^{z} \left( \ba{c} {z} \\ {n} \ea \right) 
[P^{*}_{l}(h)]^{n} [1-P^{*}_{l}(h)]^{z-n} p_{n}(h-2J)  \ee

or, using the identity $P^{*}_{l}(h)=P^{*}_{u}(h-2J)$,

\be p(h-2J)=\sum_{n=0}^{z} \left( \ba{c} {z} \\ {n} \ea \right) 
[P^{*}_{u}(h-2J)]^{n} [1-P^{*}_{u}(h-2J)]^{z-n} p_{n}(h-2J)  \ee

It follows from the above equation that the reverse trajectory will meet
the upper half of the major loop at $h^{\prime}=h-2J$ and merge with it
for $h^{\prime} < h-2J$. Thus, the task of computing the minor hysteresis
loop is reduced to range $h-2J \le h^{\prime} \le h$. We return to the
question asked at the beginning of this section. What is the probability
that a site $i$ which is up at $h$ turns down at $h^{\prime}$ ? This is
given by,

\be q^{\prime}(h^{\prime})=\sum_{n=0}^{z} \left( \ba{c} {z} \\ {n} \ea
\right)  [P^{*}_{l}(h)]^{n}[D^{*}(h^{\prime})]^{z-n}
\left[p_{n}(h)-p_{n}(h^{\prime})\right] \ee

Here $D^{*}(h^{\prime})$ is the probability that a nearest neighbor of
site $i$ turns down on the reverse trajectory before site $i$.
$D^{*}(h^{\prime})$ is determined by the equation,

\bea
D^{*}(h^{\prime})= \sum_{n=0}^{z-1} \left( \ba{c} {z-1}
\\ {n} \ea \right) [P^{*}_{l}(h)]^{n}[1 - P^{*}_{l}(h)]^{z-1-n}
\left[1-p_{n+1}(h)\right] & \nonumber \\
+\sum_{n=0}^{z-1} \left( \ba{c} {z-1} \\ {n} \ea \right)
[P^{*}_{l}(h)]^{n}[D^{*}(h^{\prime})]^{z-1-n}
\left[p_{n+1}(h) - p_{n+1}(h^{\prime})\right]
\eea

Given a site $i$ which is up at $h$, the first sum above gives the
conditional probability that a nearest neighbor of site $i$ is down at
$h^{\prime}=h$, i.e. at the very start of the reverse trajectory (and
hence remains down for $h^{\prime} < h$). The second sum takes into
account the situation that the nearest neighbor in question is up at $h$,
but turns down before site $i$ turns down on the return loop. Note that
{\em{all}} the nearest neighbors of a site i which turned up after it
turned up on the lower major loop must turn down before site $i$ turns
down on the upper minor loop.
 
The magnetization on the reverse trajectory is given by,
\be
m^{\prime}(h^{\prime})=2 [p(h)-q^{\prime}(h^{\prime})] -1
\ee

\section{Completing the Minor Loop}

We reverse the field $h^{\prime}$ to $h^{\prime\prime}$
($h^{\prime\prime} > h^{\prime}$) to trace the lower half of the
return loop. The magnetization on the lower half of the return loop
may be written as,
\be
m^{\prime\prime}(h^{\prime\prime})=2 [p(h)-q^{\prime}(h^{\prime})
+ p^{\prime\prime}(h^{\prime\prime})] -1
\ee
where $p^{\prime\prime}(h^{\prime\prime})$ is the probability that
an arbitrary site $i$ which turned up at $h$ and turned down at
$h^{\prime}$, turns up again at $h^{\prime\prime}$.

\be
p^{\prime\prime}(h^{\prime\prime})=\sum_{n=0}^{z} \left( \ba{c} {z}
\\ {n} \ea \right) [U^{*}(h^{\prime\prime})]^{n}
[D^{*}(h^{\prime})]^{z-n}
\left[p_{n}(h^{\prime\prime})-p_{n}(h^{\prime})\right]
\ee
Here $U^{*}(h^{\prime\prime})$ is the conditional probability that
a nearest neighbor of a site $i$ turns up before site $i$ turns up
on the lower return loop. It is determined by the equation,
\bea
U^{*}(h^{\prime\prime})= P^{*}(h) -\sum_{n=0}^{z-1} \left( \ba{c} {z-1}
\\ {n} \ea \right) [P^{*}_{l}(h)]^{n}[D^{*}(h^{\prime})]^{z-1-n}
[p_{n}(h)-p_{n}(h^{\prime})] & \nonumber \\
+\sum_{n=0}^{z-1} \left( \ba{c} {z-1} \\ {n} \ea \right)
[U^{*}(h^{\prime\prime})]^{n}[D^{*}(h^{\prime})]^{z-1-n}
\left[p_{n}(h^{\prime\prime}) - p_{n}(h^{\prime})\right]
\eea

The rationale behind equation (12) is similar to the one behind equation
(8). Given that a site $i$ is down at $h^{\prime}$, the first two terms
account for the probability that a nearest neighbor of site $i$ is up at
$h^{\prime\prime} \ge h^{\prime}$. Note that the neighbor in question must
have been up at $h$ in order to be up at $h^{\prime}$, and if it it is
already up at $h^{\prime}$ then it will remain up on the entire lower half
of the return loop, i.e. at $h^{\prime\prime} \ge h^{\prime}$. The third
term gives the probability that the neighboring site was down at
$h^{\prime}$, but turned up on the lower return loop before site $i$
turned up. It can be verified that the lower return loop meets the lower
major loop at $h^{\prime\prime}=h$ and merges with it for
$h^{\prime\prime} > h$, as may be expected on account of the return point
memory. The exact expressions given above have been checked against
numerical simulations of the model in selected cases resulting in
excellent agreement in all cases which were tested. 

\section{Concluding Remarks.}

The method of calculating the minor loop described above may be extended
to obtain a series of nested minor loops. The key point is that whenever
the applied field is reversed, a site $i$ may flip only after all
neighbors of site $i$ which flipped in the wake of site $i$ (on the
immediately preceding sector) have flipped back. The neighbors of site $i$
which remained firm after site $i$ flipped previously do not yield before
site $i$ has flipped again on the return loop. We have obtained above
expression for the return loop when the applied field is reversed from
$h_{ext}=h$ on the lower major loop to $h_{ext}=h^{\prime}$ ($h-2J \le
h^{\prime} \le h$), and reversed again from $h_{ext}=h^{\prime}$ to
$h_{ext}=h^{\prime\prime}$ ($h^{\prime\prime} \le h$). When the applied
field is reversed a third time from $h^{\prime\prime}$ to
$h^{\prime\prime\prime}$ ($h^{\prime\prime\prime}< h^{\prime\prime}$),
expressions for the magnetization on the nested return loop follow the
same structure as the one on the trajectory from $h$ to $h^{\prime}$. 
Qualitatively, the role of $P^{*}$ on the first leg ($h$ to $h^{\prime}$) 
is taken up by $U^{*}$ on the third leg ($h^{\prime\prime}$ to
$h^{\prime\prime\prime}$)  of the nested return loop. 

We conclude by comparing the results obtained above with numerical
simulations on Bethe lattices of coordination number $z=3$ and $z=4$, and
also contrast these results with those obtained in the one dimensional
case~\cite{shukla5}. Let us specifically choose a Gaussian distribution of
the random field with mean value zero and variance ${\sigma}^{2}$. One
generally expects the solution of an Ising model with nearest neighbor
interactions on a Bethe lattice ($z \ge 3$) to be qualitatively different
from its solution in one dimension ($z=2$), and to be similar to the
mean-field solution of an infinite-range model. However, these
expectations are not born out in the case of hysteresis in RFIM.  The
mean-field solution~\cite{sethna} does not show any hysteresis for $\sigma
\ge \sigma_{c}(\infty) = \sqrt{ \frac{2}{\pi}}$. In contrast to this,
there is hysteresis on Bethe lattices for all values of $\sigma$. 
Moreover, the behavior on lattices with $z=2$ and $z=3$ turns out to be
qualitatively similar. For lattices with $z \ge 4$, the magnetization on
each half of the hysteresis loop has a jump discontinuity for $\sigma \le
\sigma_{c}(z)$; the jump discontinuity is absent on lattices with $z \le
3$ for any finite value of $\sigma$. Figure 1 shows the major as well as a
minor hysteresis loop on a Bethe lattice with $z=3$, and $\sigma =2$. The
minor loop starts on the lower half of the major loop at $h=1.5$ and meets
the upper half at $h=-.5$ as may be expected from the theoretical
prediction.  As the magnetization on the lower half of the major loop is a
single valued function of the applied field in Figure 1, the point where
the applied field is reversed determines the minor loop uniquely.  The one
dimensional case ($z=2$) is similar~\cite{shukla5}. However, the situation
is different for $z \ge 4$. For $z=4$, we have $\sigma_{c}(4)=1.78$
approximately.  Figure 2 shows the major loop for $z=4$ and $\sigma=1.70$
with a jump discontinuity at a critical field $h_{c}$ which is slightly
higher than unity for $\sigma=1.70$. There are two values of the
magnetization at $h_{c}$ both lying on the lower half of the major loop. 
If we reverse the applied field from the value $h_{c}$, we must specify
the state of the magnetization from where the return is made giving us two
possible return loops originating at $h_{c}$. Figure 2 shows two minor
loops starting at $h=.95$ (slightly less than $h_{c}$), and $h=1.05$
(slightly greater than $h_{c}$) on the lower half of the major loop. Both
return trajectories touch the upper half of the major loop when the field
has been reversed by an amount $2J$ as expected from the theoretical
analysis. The overall agreement between the simulations and the theory is
also quite good indicating that the model considered here is
self-averaging~\cite{shukla3,shukla4,shukla5}.

\begin{figure}
\begin{center}

\includegraphics[angle=-90,width=16cm]{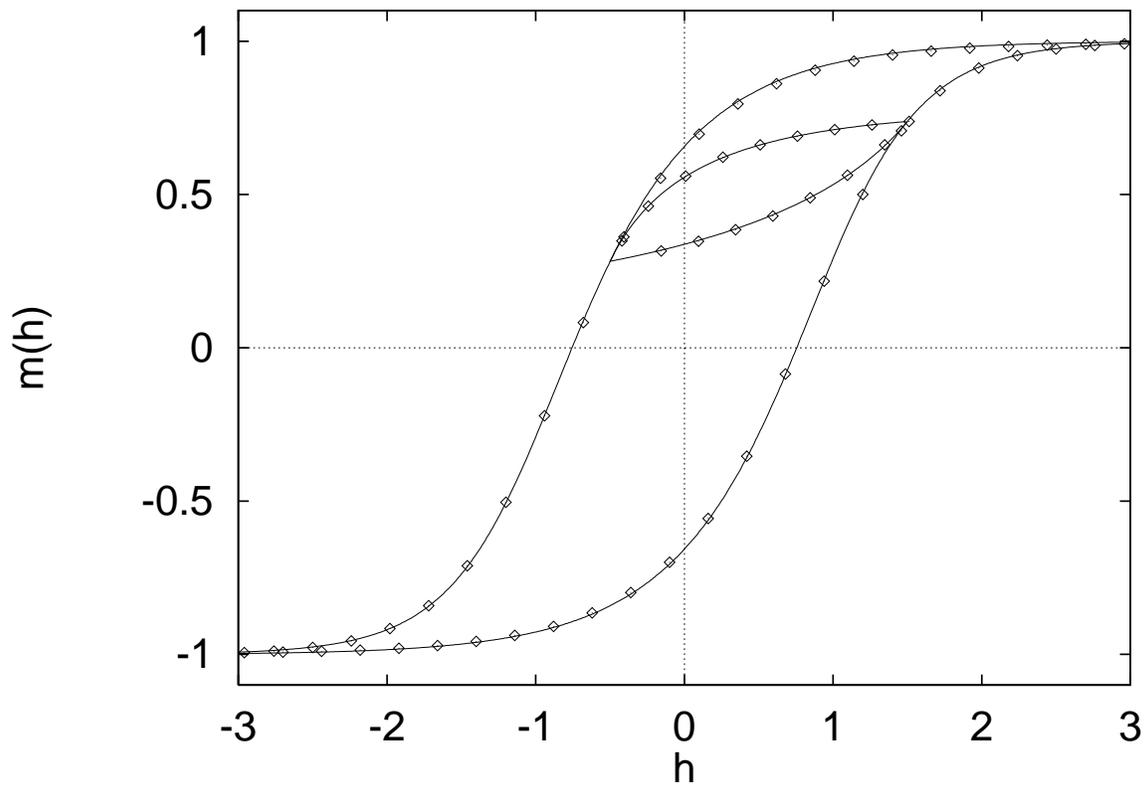}

\caption{ Major and minor hysteresis loops in RFIM ($J=1$) on a Bethe
lattice ($z=3$) for a Gaussian distribution of the random field with mean
zero and $\sigma=2$. The minor hysteresis loop is obtained by reversing
the applied field from $h=1.5$ to $h^{\prime}=-.5$ and back to $h=1.5$.
Theoretical result is shown by a continuous line, and symbols show the
data obtained from numerical simulation of the model.} 

\end{center}
\end{figure}

\begin{figure}
\begin{center}

\includegraphics[angle=-90,width=16cm]{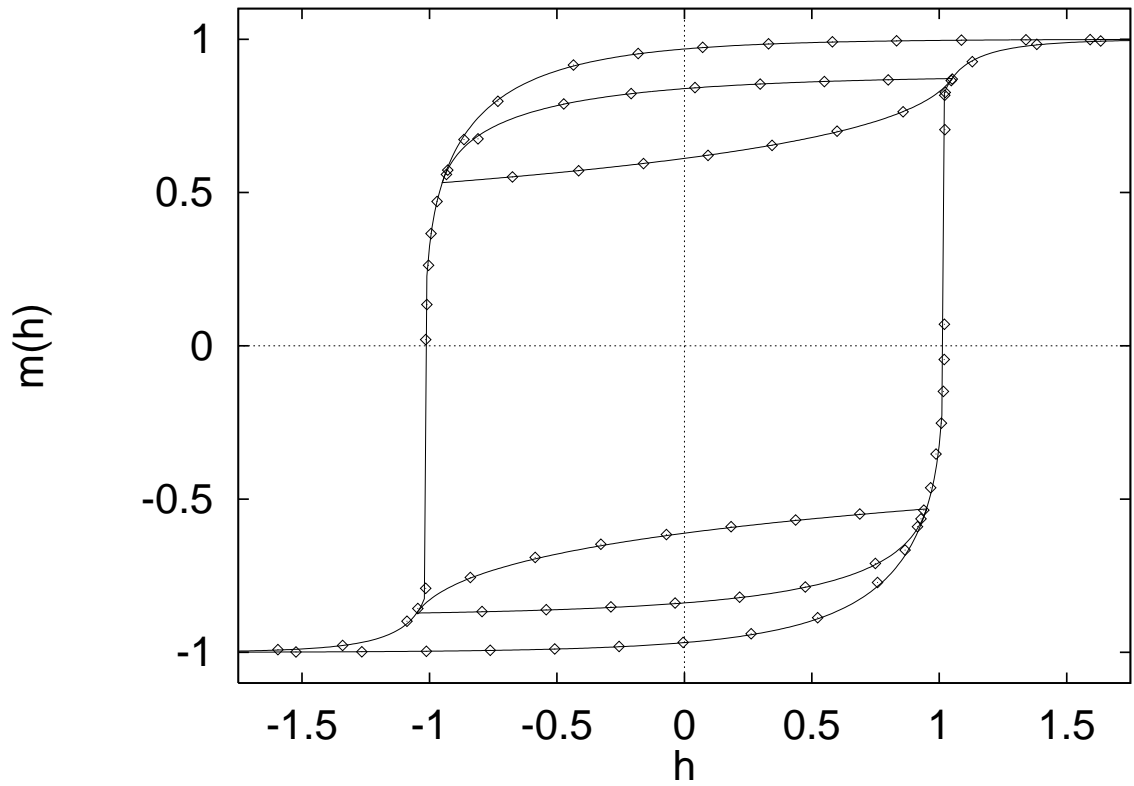}

\caption{ Hysteresis in RFIM ($J=1$) on a Bethe lattice ($z=4$) for a
Gaussian random field of mean zero and $\sigma=1.70$. Discontinuities in
the major loop vanish above $\sigma_{c}$=1.78. Two minor loops are shown
starting on the lower major loop at $h=.95$ and $h=1.05$ respectively. As
in Figure 1, theoretical result is shown by a continuous line, and symbols
show the data obtained from numerical simulation of the model.  Note that
the minor loops touch the upper major loop when the applied field has been
reversed by an amount $2J$.}

\end{center}
\end{figure}


\begin{thebibliography}{99}

\bibitem{sethna} J P Sethna, K A Dahmen, S Kartha, J A Krumhansl, B W
Roberts, and J D Shore, Phys Rev Lett 70, 3347 (1993). 

\bibitem{shukla1} P Shukla, Physica A 233, 242 (1996); P Shukla, R Roy,
and E Ray, Physica A 275, 380 (2000); P Shukla, R Roy, and E Ray, Physica
A 276, 365 (2000). 

\bibitem{shukla2}P Shukla, Physica A 233, 235 (1996). 

\bibitem{shukla3}D Dhar, P Shukla, and J P Sethna, J Phys A: Math. Gen. 
30, 5259 (1997).

\bibitem{shukla4} S Sabhapandit, P Shukla, and D Dhar, J Stat Phys 98, 103
(2000).

\bibitem{shukla5} Prabodh Shukla, Phys Rev E 62, 4725 (2000). 

\end{thebibliography}
\end{document}